# The Shady Light of Art Automation

**Dejan Grba**

Artist, researcher, and scholar
Belgrade, Serbia
dejan.grba@gmail.com

**Abstract**

Generative artificial intelligence (generative AI) has entered the mainstream culture and become a subject of extensive academic investigation. However, the background and character of its impact on art require subtler scrutiny and more nuanced contextualization. This paper summarizes a broader study of the roles that AI's conceptual and ideological substrata play in influencing art notions. The focus is on divergent and often questionable but somehow coalescing ideas, values, and political views in the computer science and AI/tech industry that generative AI and other art-related AI technologies propagate to contemporary art and culture. The paper traces the main strands of this complex transference and concisely critiques their key aspects.

## Introduction

The advent of large-scale models, also known as foundation models, [1] diffusion models, [2] and multimodal generative models [3] between 2021 and 2022 led to notable performance improvements and scope extension in generative artificial intelligence (generative AI). The expansive deployment and rapid commercialization of multimodal generative models, for instance in various media-generating applications since 2023,[1] have made generative AI a part of the mainstream culture. Generative AI's applications in artmaking and associations with creativity have been subjected to an extensive academic investigation.[2]

While AI-induced changes in art's expressive realm (new topics, techniques, and forms), scholarship (applications in art history, museology, and digital humanities), and socioeconomic status (modalities of production and monetization) have been extensively explored,[3] the character and background of AI's impact on art notions require a deeper and subtler examination. Taking generative AI as a context, this paper encapsulates the key points of a broader study about AI's conceptual and ideological undercurrents that impact art notions. We sketch a flux of diverse but often questionable technical ideas, ethical values, and political views that generative AI and other art-related AI technologies transmit from the computer science and AI/tech industry to contemporary art and culture. Focusing on the main areas of this complex realm of influence, we outline their aspects in the following two sections and, in conclusion, provide a concise critique of their manifestations in generative AI. While focusing on sociopolitical, economic, and ethical implications for artmaking, we recognize that generative AI's underlying computational principles warrant exploration as a potentially distinct paradigm for creative expression unburdened with the imperative to imitate human creativity. Consequently, we believe that the artistic sense of generative AI goes beyond the mimicry of human features and wish to emphasize that the conceptual conflation of AI with human intelligence and artistic motivation is shortsighted and may be harmful. In this context, however, we do not elaborate on how (generative) AI could challenge or expand our definitions of creativity and authorship, which has already been done, notably by [11] and [12].

Tracing the instrumentalization of art and creativity as the cultural normalizers of AI and the tech industry's dubious values is a demanding task, and it is important to recognize its limitations. To encapsulate a significantly larger study in this paper, we map sweeping and convoluted subjects from diverse disciplines into a compact narrative, so certain issues and their interrelations are omitted or not discussed in detail. We summarize the trends of AI's notional influence on art rather than discussing specific artworks that exemplify it, which would require a separate publication of considerable volume. Generative AI issues extend the comprehensive set of requirements for responsible artists, educators, and policymakers to tackle the challenges posed by AI (see [13]). Articulating these requirements with precision and nuance is the objective for future work. Several other indicated topics warrant further examination and dedicated publications, such as how AI's ideological undercurrents influence artistic production by designing and profiling AI tools marketed to artists. Another important topic concerns the factors for blending individually incongruent ideas and tendencies in the computer science and AI industry and the mechanisms of their cultural impact in areas besides artmaking. Taking these constraints into account, we offer a critical perspective for looking at the intersections of AI and art, which may be conducive to cultivating an informed and responsible approach to the contemporary AI-influenced society.

---

[1] Popular examples include Midjourney, Inc.'s Midjourney, OpenAI's DALL·E and Sora, Stability AI's Stable Diffusion, Meta's Make-A-Video, Runway's Gen-2, and Google's Lumiere.

[2] For an overview of commonly addressed aspects, see [4] and [5].
[3] See [6], [7], [8], [9], and [10].

## AI and Art Notions

The art- and creativity-related AI technologies, economies, and discourses introduce assumptions, views, and generalizations about art and creativity, which directly or indirectly affect artistic practices, art notions, and speculations about the future of artmaking. [14] These changes unfold within a multilayered sphere of influence whose main components are: the ontological and taxonomic aspects of art notions (identification), the normalization of AI-powered artmaking, the representation of AI art, and the promotion of AI as a means to disseminate or democratize creative expression.

### Identification

The cultural, economic, and political circumstances of making and experiencing art constantly change, so the notions of what an artwork is evade stable and consensual forms across time periods, geographies, and societal domains. [15] Artists affect them by modifying or breaking up with the established professional norms, theoretical views, and cultural traditions, which in turn catalyze notional changes in art's identities and artists' roles. For instance, Marcel Duchamp's transposition of artmaking from the reconfiguration of matter into a cognitive process of relational creativity and discovery in the early 20$^{th}$ century has substantially driven art's accentual shift from formal representation to a conceptual exploration that equally favors natural, artificial, physical, and imagined elements. [16][17] Since then, the conceptual frameworks and vocabularies for dealing with the creative approaches in the emerging arts have been incessantly created, modified, and discarded.

Consequently, art appreciation has become open for objects, events, or processes that do not need to be aesthetically pleasing if they facilitate meaningful communication, discovery, and learning. [18] It requires an experientially, intellectually, and emotionally competent spectatorship attuned to the artworks' demands and depends on the knowledge and understanding of art's historical and contemporary dynamic. These demands are not always met in pondering art's "essences". The attempts at formulating universal art-identification algorithms in philosophy and art theory chronically fail to tackle art's open-endedness. [19] In the informal discourse, the entitlement for determining and judging artworks by "individual taste" without revealing its informative qualities often shields uneven art knowledge or comprehension.

Missing or ignoring the import of art's evolution and its emancipatory implications for art's meaning leads to misconceptions of established as well as contemporary art fields with solid historical backgrounds, such as AI art. Besides the term "AI art", which is commonly used thanks to its inclusivity for all relevant types of expressive approaches contingent on AI's cultural contexts, the scholarly and popular discourses brim with potentially misleading names that belong to older and broader art fields, such as "digital art", "computational art", and "generative art", or constraining labels such as "machine learning art". The hyped-up trends in AI tech push the online, media, and pundit terminologies toward historically myopic exclusivity, so AI art is sometimes conflated only with practices that utilize machine learning [20] or generative AI technologies, [21] or just with AI-generated visuals. [22] In AI's multifaceted cultural sway, taxonomic and narrative affinity for the creative uses of vogue technologies reduce the space for appreciation of complex art fields to market labels and promote their uncritical appreciation.

### Normalization

The popularization of AI technologies for expressive purposes and the introduction of consumer-grade AI products designed specifically for artmaking have never been just casual byproducts of AI development. They belong to the history of corporate instrumentalization of art and creativity to culturalize computer technologies, which began in the late 1950s. [23] The preceding AI industry's stabs with machine learning include style transfer apps/services (since 2014), DeepDream (since 2014), ArtBreeder (since 2018), Runway (since 2018), and art support/showcase programs such as Google's Artists and Machine Intelligence (since 2016).[4] [25] Pitching AI applications for artmaking benefits both the AI industry's marketing and development as widely adopted software tools become "indispensable", their usage provides beta testing, feedback, and learning data from a large user base, and helps associate AI with unique human faculties such as artmaking.

These trends combined with the improved realistic rendering and multimodality of new learning models and techniques to make generative AI art a social phenomenon that spawned a lively economy by leveraging the blockchain-enabled tradability of digital artefacts. [26] Like in the initial boom of text-to-image production, the most prevalent and lucrative artmaking practices with generative AI privilege figurative (descriptive) plastic motifs in popular genres of "surreal" or fantasy art, game design, comics, anime, or illustration, with a fixation on surface aesthetics and stylistic norms at the expense of other poetic factors. [27] Their commercialization encourages the production and consumption of desirable (aesthetically pleasing and "collectible") digital artefacts, which may resurge the simplistic notions of art.

### Representation

Malleable concepts such as agency, authorship, and originality were among the central topics of modernism and computational artists have addressed them since the 1960s, but the specter of misidentified creative autonomy has always been haunting their efforts. That is because sophisticated computing technologies can easily trick us into conflating the ultimately essential human creativity with its highly formalized and heteronomous simulations. The anthropomorphic legacy has been endemic in AI art, from Harold Cohen's phraseology about his painting/drawing robot AARON (1971–2016) [28] to contemporary artists' claims about the creative agency of machine learning programs (see, for example, [29] and [30]).

---

[4] They were preceded by an academic project—a machine learning software for artists named The Painting Fool that Simon Colton and the Computational Creativity Group at Imperial College in London developed in 2012. [24]

The art market's unclear rhetoric about the authorship in AI art has been commercially motivated, as the insinuations of artworks produced by expressively motivated AI systems (rather than by humans who deal with AI technologies) cash in on the momentum of the ongoing AI hype. [31][32] This boosts artists' "immunity" to the criticism and debunking of anthropomorphic notions (see [33]), and they frequently discuss their AI devices as "creative collaborators", "partners", or "companions". [34] The audience's romantically skewed (anthropocentric) perception and virtue signaling about AI artworks reinforce the myths of machinic agency, which in turn encourages artists and the art market to exploit them further. [35][36] Intentionally or not, some AI studies join this feedback loop by metaphorizing generative AI media synthesis processes as forms of the artists' externalized visual cognition [37] and identifying generative models as co-creative agents. [38]

Similarly, the critical scholarly views of generative AI media as artworks whose authorship is undefinable or distributed among AI scientists and programmers, and the views that recognize generative AI models as artworks but not the media they produce (see [39]) reshuffle the contestation points about authorship that engulfed AI art as soon as it gained cultural prominence in the late 2010s. A more coherent take on generative AI regards it as a sophisticated remediation apparatus related to earlier remix techniques because generative models depend on a predictive amalgamation of samples of artefacts whose preestablished styles and other features get reflected in the output media. [40][41]

## Democratization

Since Alan Kay and Adele Goldberg's influential concept of the computer as an "active metamedium" in 1977, [42] several waves of digital creativity enhancements have brought about novel tools and methods for content creation and reinvigorated older ones (such as bricolage and remix), with convergent cultural consequences.[5] Regardless of their accessibility levels—from user-friendly apps to programming languages with steep learning curves—computational artmaking tools translate certain features of previously established artistic media or techniques and often open new creative possibilities. They usually allow programmability or extensibility but feature different motivations and affects that inevitably miss some subtleties of technical decision-making inherent to their source art practices. The expressive routes, conceptual values, and aesthetics of digital media tools are further influenced by the tech designers' (rarely neutral) ideas about their sense and purposes and the production standards/frameworks, trade-offs, and compromises in research and development processes. Finally, the producers' legislative concerns, economic interests, and political views shape the interface metaphors and operational protocols of the end products. [43][44]

These technical, ideological, and economic impositions on digital media may still stimulate sophisticated creative thinking when they are recognized and acknowledged. But early adopters are usually more preoccupied with meeting the new tools' cognitive demands and exploring their exciting features than studying their art-historical and techno-cultural backgrounds, which leads to formalism and technocentrism. [45] Once digital tools are widely adopted, their expressive conditioning remains mostly hidden behind typical usage scenarios and conventional practices that predispose trivial or uncritical expression. [46]

By dissolving the previous distinctions between professional and nonprofessional content producers, the sophistication of digital media has also contributed to the diffusion of the artworld throughout the 20[th] century. It became a hybrid, multicultural conglomerate of ideas and approaches, with many centers and peripheries where artists often evade codified roles and engage in revisiting and mixing genres and styles. Since the introduction of blockchain technologies in the late 2000s, these changes have encouraged various types of creative hobbyists, prosumers, amateurs, and weekend artists to compete in an attention economy that often turns them into commodities. [47] After integrating blockchain technologies in the late 2010s, the art market has gradually started sharing its selection criteria with crypto investors acting as art collectors and with monetization algorithms on NFT trading websites (see [48] and [49]). This confluence of creative directions and economic incentives has fomented a straightforward, mostly automated proliferation of tradeable eye-catching digital artefacts whose styles converged toward derivative, platform-powered aesthetics akin to zombie formalism. [50][51] Boasting a comparable perpetuation of cultural norms, stereotypes, biases, and hegemonies, the expressive trends in generative AI media production serve as reminders that art's notional open-endedness equally applies to kitsch as "art's shammy doppelgänger" and that their affair has now become more intense and promiscuous.

With the abundance of accessible and easy-to-use tools that "bring the AI power to the masses by allowing just about anyone" to become an artist, [52][53] generative AI is broadly presented as a technology that democratizes artmaking. However, such claims disregard that artmaking is not merely a matter of access to the means of expression and presentation.[6] On the contrary, generative AI may destabilize and potentially kill off many skilled jobs in the arts. Faster development with fewer staff appeals to illustration, design, gaming, film, and other creative enterprises where profitmaking takes primacy over the concerns about their human resources' wellbeing. While automation can be useful for relieving tedious, dehumanizing, and hazardous types of labor, [54] many forms of creative work emulated and supplanted by generative AI are on the opposite pole of these categories and some authors argue that the core logic of AI development goes against the respect for human's intellectual and economic rights. [55]

---

[5] The most notable enhancements of digital content creation came with the introduction of the personal computer in the 1980s, Apple's doctrine of user-friendly computation in the mid-1980s, the Internet in the 1990s, Web 2.0 technologies in the mid-2000s, blockchain in the late 2000s, and the NFTs and AI media generators since the mid-2010s.

## The Undertows

AI-influenced changes in art notions, artmaking practices, and other forms of creative expression transmit philosophical premises, technical concepts, political views, and ideological tendencies from AI science, technology, and industry to the mainstream culture and broader society. Many forces in this diverse but coalescing flux of influences have the overtones of alienation, sociopathy, and misanthropy, which largely escape the debates about AI's cultural effects.

### The Computer/Human Equation

The epistemological and metaphysical confusions caused by conflating human features with machine performance have rendered AI inseparable from anthropomorphism. This innate psychological tendency to assign human cognitive traits, emotions, intentions, and behaviors to nonhuman entities or phenomena [56] permeates the foundational concepts of intelligence and terminologies in AI science and industry as well as popular discourse (see, for instance, [57] and [58]). The performance of state-of-the-art AI systems is frequently associated with human cognitive traits such as causal modeling, active social learning, conceptualization, subconscious abstraction, generalization, analogy-making, and common-sense reasoning—the very capabilities they lack the most. [59] Such associations can have profound consequences in high-risk and other sensitive domains: AI design and deployment unfold within a socially constructed context in which humans deliberately outsource certain tasks to machines, but anthropomorphism implicitly grants "them" (machines) a degree of agency that overstates their true abilities. Crucially, the transfer of operational authority to algorithms does not absolve humans of responsibility. [60]

The awkward and paradoxical tendencies toward the understanding of computers vis-à-vis humans reach back to the foundations of modern computer science. One of the unfortunate consequences of Alan Turing's legacy is the intentional or accidental provision of a "scientific basis" for the mutual equalization of human beings and computers. Turing ostensibly went from analogizing the isolated features between human and machine computation in the concept of Turing machine ("On Computable Numbers" 1936) [61] toward conflating human beings with computing machines in his later work. In a 1950 paper "Computing Machinery and Intelligence", [62] Turing proposed the Imitation Game (Turing Test) as a method for testing a computational machine's ability to exhibit intelligent behavior equivalent to, or indistinguishable from, a human. However, the proposal was centered around an unclear concept of intelligence and left many other parts of the discussion open to interpretation, causing a long-lasting controversy. [63]

Turing's critics claim that he devised the Imitation Game to legitimize the "null hypothesis" of no behavioral difference between certain machines and humans, which was arrogant because it assumed understanding human cognition without first obtaining a firm grasp of its basic principles. [64][65] Turing's affinity for the "null hypothesis" also provides grounds for an argument that psychological issues evident throughout his life elicited a misanthropic bitterness, which motivated the computer-human analogy.[7] Beyond Turing, the AI community's leniency toward the persistence, range, and character of its members' quirks and grotesque notions that signal both conceptual and mental issues [67] indicate the abuse of the reasonable need to tolerate the intellectual challenges and mental tolls of cutting-edge thinking, which is regressive and irresponsible. These unresolved ambiguities translate to the reality in which the rapid industrialization and widespread application of AI technologies bring about the concentration of wealth and political power that leads to a society contingent on corporate AI interests.

### Datafication and Fauxtomation

Data collection and quantization, behavioral tracking, predictive modeling, and decision-making manipulation have long been essential strategies for large-scale information-dependent institutions such as governments, industry, marketing, finance, insurance, media, and advertising. By coupling massive data capture with sophisticated statistical algorithms, modern AI increases the extent, intricacy, and efficacy of the social engineering strategies in both companies and bureaucratic systems traditionally mandated to impose goal-oriented rational ordering on increasingly complex societies and pressured to deliver optimized, metrics-based management and governance. [68][69][70]

Thus, the same features of abstraction and optimization that make AI so useful can also reinforce bureaucratic callousness in social profiling and targeting as the fact that diversity, variety, and complexity of experience surpass and

---

[6] If we broadly understand democracy as a system for configuring and operating governance mechanisms of human associations according to the will and best interests of their constituents, the idea of democratizing artistic expression makes little sense in general. Artmaking is predominantly an individualistic enterprise and the artworld and art market are inherently competitive and often adversarial, not democratic institutions.

[7] Henry O'Connell and Michael Fitzgerald's analysis of Turing's biography and contemporaneous accounts [66] concludes that he met Gillberg, ICD-10, and DSM-IV criteria for Asperger's syndrome, which places him within the autism spectrum disorder.

therefore resist abstract representations often gets lost or overlooked in the practice. [71] The widespread institutional and business adoption of AI increases the scope of bureaucratic operations that can be algorithmized, but opaque AI decision-making processes also expand the scope of unaccountable policymaking and administrative violence. For instance, it can facilitate the incorporation of existing prejudices that amplify the injustices of welfare systems or the development of exploitative algorithmic frameworks for productivity extraction/maximization, and workforce concealment. The AI industry demonstrates this by frequently applying abusive labor policies and unethical human resources management that evade control and regulation although they have been documented repeatedly and thoroughly. [72][73] Generative AI's development demands for human labor have somewhat changed from their role in trailblazing the preceding AI techniques and pipelines [74] but remain high and exploitative. Building foundational models requires massive human-judgment-based work on data after pre-training stages, also called the "post-training alignment". It includes numerous repetitive and meaningless tasks that outsourced online crowd-workers in the global South, workers in start-up platforms, and the base workforce stack of the AI industry fulfill, often under inadequate, precarious, and surveilled conditions. [75][76]

The concealment of AI's social rootage, the filtering of human benefits from using AI technologies, and the misrepresentation of interests in the social conflicts they foment[8] are important factors of modern AI's economic and political power. [77][78] In aggregate, these trends and practices contrive an illusion that human-created and human-dependent AI systems have high levels of material abstraction and functional autonomy,[9] which exacerbates structural violence and induces learned helplessness among its victims. [83]

**The Ideological Assemblage**

Exploitative anthropomorphization, elaborate illusionism of human-independent automation, and algorithmic labor maximization are historically ingrained in corporate AI's social politics. The biopolitical paradigm of cybernetics in the 1950s, which paved the way for modern computer science and industry, already placed humans in an ambiguously symbiotic relationship with machines and was ideologically geared at subjecting society to technology. [84][85] Since the mid-1960s, the worldviews in computer science communities and IT industries, particularly in the US, have been shaped by a bizarre ideological conglomerate of doctrines, such as techno-utopianism, counterculture, individualism, libertarianism, and neoliberal economics. [86][87] This assemblage, also called the Californian ideology [88] and cyberlibertarianism, [89] gained momentum during the tech boom of the 1990s. It comprises ideas fueled by the zeal for technologically mediated lifestyles and future visions steeped in libertarian notions of freedom, social life, and economics.[10] It endorses technological determinism [91] and promotes techno-solutionism[11], radical individualism, deregulated market economy, trust in the power of business, and disdain for the role of government. [94][95] These values fully make sense only within the context of the right-wing political milieu and are often spiked with pseudo-philosophical rhetoric, such as Objectivism. [96][97]

After the introduction of blockchain technologies, the cyberlibertarian techno-solutionist politics has been radicalized by the burgeoning start-up community of predominantly white male entrepreneurs obsessed with quick success and tending toward sexism, racism, misogyny, homophobia, and transphobia. [98] Most early cryptocurrency developers and influential adopters belong to intersecting movements and groups such as cypherpunks, crypto-anarchists, transhumanists, Singularitarians, Extropians, self-described hackers, open-source software developers, and tech-savvy entrepreneurs. Leading crypto and AI investors, such as Elon Musk, Peter Thiel, Eric Raymond, Jimmy Wales, Eric Schmidt, Saifedean Ammous, and Travis Kalanick, openly adhere to libertarianism and Objectivism, and promote economic views that range from the Austrian and Chicago schools of economics to the Federal Reserve conspiracy theories. [99] Therefore, despite the nominal commitment to widely acceptable social values, many crypto-economic and IT ventures epitomize right-wing politics.[12] This does not mean that entire tech and AI industries are right-wing, but that a web of right-wing politics and reactionary opinions spreads across them, often by actors in pivotal positions. Behind the facade of objectivity, rationality, progress, and political correctness, the AI industry's reality is dominated by aggressive competitiveness within an adversarial business culture that reflects the most unpardonable tenet of capitalism: prioritizing profit over people. [101]

Generative AI is part of the continuous alignment of mainstream AI with info-capitalist ideological and socioeconomic frameworks. It can be seen as the forefront of the reiterative entrepreneurial process toward emancipating capital from humanity, in which human labor and data provision are exploited to build systems that automate certain tasks and reconfigure humans' roles for the next iteration. [102] Some authors deem such logic morally untenable and destructive, [103] others remain undecided [104] or claim the opposite. [105] Regardless of our positions on this spectrum of views, it is worth remembering that insofar as we take advantage of AI's sociotechnical regime, we share a degree of responsibility for its existence and consequences.

---

[8] Debates about universal basic income (UBI) are another instance of corporate AI's hypocrisy. They mostly revolve around AI-caused job clearance instead of acknowledging that the welfare state instruments such as UBI are necessitated by the inequality of wealth generation and distribution whose long history reaches far behind AI and computing technologies.

[9] See [79], [80], [81], and [82].

[10] See also [90].

[11] Techno-solutionism is a tendency to favor advanced technologies for solving complex widely relevant problems instead of making attempts at addressing their—often crucial—structural causes. Foregrounding technical fixes also covers up these underlying causes. See [92] and [93].

[12] For additional examples of the AI industry's ties to right-leaning political ideologies, see [100].

## Conclusion

While these and other issues of computer science and AI have been examined in science and technology studies, they require proper articulation and wider attention in artistic communities. Generative AI is the latest instance of the AI industry's utilization of art and creative expression as avenues to culturalize its products, secure its economic interests, and promote its political views. The problematic underpinnings of this influence are largely obscured or attenuated in the debates about AI's cultural impact and remain underexposed in critical AI scholarship. Nevertheless, they hijack our cultural intuition, [106] translate into art practices and their public reception, [107][108] channel the professional and popular art discourse, influence the notions of art and creativity, and shape the visions of the future and their actualizations. [109]

More specifically, the promotion of generative AI as an art-democratizing technology refrains the cyberlibertarian myths about the democratizing powers of markets and digital technologies [110] and supports the info-capitalist creativity imperative. [111][112] It cynically exploits false claims about limited or worn-out human creativity while simultaneously celebrating AI developers' creative capacities for building creativity-boosting software tools.[13] The insinuations of generative AI models' autonomous artistic faculty exploit the evolved human bias toward detecting agency. They simultaneously rely upon the concealed heteronomy of digital computation technologies and dismiss the centrality of sociocultural relations for artmaking. [114] Motives for relegating expressively relevant decisions to AI systems often converge into hedging artistic responsibilities and foregrounding the benefits of automated cultural production. [115] Artists' compliance with the censorship criteria imposed on generative models by their owners [116] upholds the AI industry's control of clients' socioeconomic benefits from leveraging its products. At the same time, artists' pragmatic adoption of first-aid tech tools against the misappropriation of their work for generative model training, such as data poisoning or style masking, sustains their creative and economic dependence on the AI industry, research, and academia. It inadvertently stokes enthusiasm about AI's potential for transforming art, which plays in tune with the techno-solutionist rhetoric whereby only the tech (but not the regulation of techno-economic power) can save us. [117]

In a broader perspective, we should address the shady undertows of AI-influenced culture more assertively as amalgamations of historically entrenched economic interests, [118] self-serving anthropocentrism, [119] the human propensity for deception and self-deception, [120] and exploitative virtue signaling. [121] The avoidance or euphemistic treatment of unflattering but costly human traits that drive these undertows is partly caused by the false dichotomy of culture versus biology. Nevertheless, culture rests upon and emerges from our evolved mental architecture and cannot be understood without it. [122] A sincere reassessment of this architecture should drive the fundamental sociocultural changes we claim to be seeking so much because AI science and industry's impact on our lives is not a magical effect of some cosmic teleology, but the cumulative outcome of human motives and actions. It would be reckless to relinquish it to an exclusive social sector with a proclivity for relational deficiencies, psychological disorders, and abusive ideologies. Instead, it is crucial to curtail the absurd fiction of its inevitability, devise instruments for its meaningful critique, and attain the political will to assume its control.

---

[13] See, for instance, [113].

## Author Biography

Dejan Grba is an artist, researcher, and scholar who explores the cognitive, technical, expressive, and relational aspects of emerging media arts. He has exhibited in the Americas, Europe, Asia, and Australia, and published papers in journals, conference proceedings, and books worldwide.